\documentclass[aps,prl,reprint,showpacs,floatfix]{revtex4-1}
\usepackage{amssymb,amsmath,graphics,epsfig}

\newcommand{\rem}[1]{ }
\newcommand{\beq}{\begin{equation}}
\newcommand{\eeq}{\end{equation}}
\newcommand{\bea}{\begin{eqnarray}}
\newcommand{\eea}{\end{eqnarray}}

\newcommand{\ket}[1]{\left| #1\right>}

\def\aj{{\it Astron. J.}}
\def\apj{{\it Astrophys. J.}}
\def\apjl{{\it Astrophys. J. Lett.}}
\def\apjs{{\it Astrophys. J. Suppl.}}

\def\mnras{{\it Montly Not. R. Astron. Soc.}}

\def\nat{{\it Nature}}
\def\prl{{\it Phys. Rev. Lett.}}

\def\prd{{\it Phys. Rev. D}}

\def\physrep{{\it Phys. Rep.}}

\def\jpha{{\it J. Phys. A: Math. General}}
\def\aa{{\it Adv. in Astron.}}

\def\mpl{{\it Mod. Phys. Lett.}}
\def\physatomnucl{{\it Phys. Atom. Nucl.}}

\def\jcap{{\it J. Cosmol. Astropart. Phys.}}
\def\adp{{\it Annalen der Physik}}
\def\jhep{{\it J. High Energy Phys.}}
\def\sovjnp{{\it Sov. J. Nucl. Phys.}}

\begin{document}

\title{Cosmological Simulations of Multi-Component Cold Dark Matter}
\author{Mikhail V. \surname{Medvedev}} 
\affiliation{Institute for Theory and Computation, Harvard University, 60 Garden St., Cambridge, MA 02138}
\affiliation{Department of Physics and Astronomy, University of Kansas, Lawrence, KS 66045}
\affiliation{ITP, NRC ``Kurchatov Institute", Moscow 123182, Russia}

\begin{abstract}
The nature of dark matter is unknown. A number of dark matter candidates are quantum flavor-mixed particles but this property has never been accounted for in cosmology. Here we explore this possibility from the first principles via extensive $N$-body cosmological simulations and demonstrate that the two-component dark matter model agrees with observational data at all scales. Substantial reduction of substructure and flattening of density profiles in the centers of dark matter halos found in simulations can simultaneously resolve several outstanding puzzles of modern cosmology. The model shares the ``why now?'' fine-tuning caveat pertinent to all self-interacting models. Predictions for direct and indirect detection dark matter experiments are made.
\end{abstract}

\pacs{95.35.+d, 95.30.Cq, 98.80.-k, 14.80.-j}
\maketitle

{\it Introduction.} --- Dark matter (DM) constitutes about 80\% of matter and 25\% of the total energy density in the universe but its nature remains completely unknown. The existence of DM requires revision of the present day physics. Most likely, DM is a hypothetical particle or particles beyond the standard model \cite{DMcandidates-rev}. 

The current heuristic paradigm of the cold dark matter with a cosmological constant ($\Lambda$CDM) is remarkably successful at reproducing the large-scale structure of the universe but appears to disagree with observations at small scales. First, simulations predict the overabundance of small mass (dwarf) halos as compared to the much lower number of the observed satellite galaxies in the Local Group \citep{Moore+99, Klypin+99, Kravtsov10, Zwaan+10} and in the field as inferred from the ALFALFA survey \citep{ALFALFA}. This problem was termed the ``substructure problem'' or ``missing satellite problem.'' Second, the cuspy $\rho\propto r^{-1}$ DM density profiles found in $\Lambda$CDM simulations \citep{NFW97} disagree with the rotation curves of dwarf and low surface brightness galaxies, which indicate flattened or cored density profiles \citep{SB00, Gentle+04, Salucci+07, Donato+09, deBlok+08, deBlok10, KdNK11}. Observations of galaxy clusters also indicate the presence of cores \citep{DMinClusters09}. Moreover, the largest $\Lambda$CDM subhalos in the Local Group-type environments are too dense in their centers to host any of the dwarf spheroidal galaxies around the Milky Way and Andromeda galaxies, and in the field \citep{B-K+11, B-K+12, P+14}. These two, perhaps related, problems are known as the ``core/cusp problem'' and ``too-big-to-fail problem'', respectively. Numerous attempts to reconcile the $\Lambda$CDM model with observations using baryonic processes made so far (modified star formation, tidal gas stripping, supernova feedback) are inconclusive \citep{Kravtsov10, Govenato+10, DMinDwarfs11, DMinDwarfs12, Penarubia+12, G-K+13,G-K+14}. This is because the latter problems require strong feedback and, hence, larger star formation whereas the substructure problem requires just the opposite -- the suppressed star formation. 
Contrary to the early expectations, a mild DM paradigm shift to adopt warm dark matter (WDM) \citep{A-R+01, Bode+01, Schneider+14} also fails to resolve all these problems altogether \citep{KdN+10, V-N+D11, Maccio+12} due to the similar constraints on the DM particle mass (but see Refs. \citep{DK84, DKK89, BK90, A+12, B+12, A+13, B+13, C+13, Bh+13} for hybrid models). 

Inability of conventional physics to resolve the  aforementioned problems within the collisionless CDM paradigm can indicate that DM may exhibit non-gravitational properties as well. The most natural alternative is to admit a large interaction cross-section of DM with itself \citep{CMH92, SS00} but not with normal matter. Contrary to the early claims \citep{Burkert00, Yoshida+00a, Dave+01}, such self-interacting dark matter (SIDM) was successful to explain the origin of cores without violating any constraints on the velocity-dependent cross-section $\sigma(v)$ \citep{Colin+02, +Shapiro+05, bullet08, A-H+09, F+10, LW11, VZL12, ZVW13}; however it completely fails to solve the substructure problem \citep{VZL12}. Interestingly, SIDM can naturally explain the presence of supermassive black holes in red bulgeless galaxies \citep{Satyapal+14} and their very early formation \citep{Willott+07, Treister+13} via gravitational collapse of the central (collisional) parts of DM halos \citep{O00, HO02} --- the process that is absent in the ``vanilla CDM'' paradigm. At last, the existence of a narrow plane of the Andromeda dwarf satellites \citep{Ibata+13}, which has no explanation within collisionless CDM, can potentially be addressed in SIDM, because collisionality induces viscous drag on subhalos (whether it is enough is unknown).

However, an important possibility that some DM candidates are quantum-flavor-mixed particles, e.g., a neutralino, an axion, a sterile neutrino, has not been considered so far. In this Letter we demonstrate from the first principles via $N$-body cosmological simulations that even the simplest model with two-component quantum-mixed DM with small mass-degeneracy agrees with observational data at both large and small scales, thus may be settling the above problems altogether. Moreover, it also agrees with the observational constraints on $\sigma(v)$ set by SIDM models \citep{bullet08, A-H+09, F+10, LW11, VZL12, ZVW13}. At last, the model makes predictions for and is testable with direct and indirect detection DM experiments. 

{\it Model.} --- First, we postulate that the dark matter particles are flavor-mixed. Generally, a mixed particle of flavor $\alpha$ is a superposition of several mass-eigenstates $\ket{f_\alpha}=a_1\ket{m_1}+a_2\ket{m_2}+\dots$, where $\ket{f}$ and $\ket{m}$ denote wave-functions being flavor and mass eigenstates, and $a_1,\ a_2,\dots$ are the elements of a unitary matrix. Here we consider the simplest DM model with two flavors and two mass eigenstates only \citep{M10, M14}, i.e., the two-component DM (2cDM) model. The masses of the mass eigenstates are $m_h$ and $m_l<m_h$, i.e., `heavy' and `light'. Generally, $\ket{m}$'s have different velocities \citep{AS09, Giunti01} and propagate along different geodesics. Hence, they can be spatially separated by gravity during structure formation: the eigenstates with smaller speeds become trapped in a growing halo earlier than the faster ones. The DM halos are, thus, self-gravitating ensembles of non-overlapping wave-packets of heavy and light eigenstates. 

Second, we postulate that DM particles can interact with each other non-gravitationally with some velocity-dependent cross-section, $\sigma(v)$, which is consistent with the existing SIDM constraints. It is customary in cosmology to parameterize it as $\sigma(v)=\sigma \left(v/v_0\right)^{-a}$, where $\sigma$ and $a$ are parameters and $v_0$ is a normalization constant. Previous studies and observational data allow for $a\gtrsim1$ \citep{Colin+02, HO02, bullet08}, so $a=1$ is used in the simulations reported here. This $1/v$-dependence is also natural for mass-eigenstate conversions \cite{M14}. Observations constrain the ratio $\sigma/m$, where $m$ is the DM particle mass, to be in the range $0.1\lesssim\sigma/m\lesssim{\cal O}(1)$~cm$^2$/g for the assumed normalization $v_0=100$~km/s \citep{Colin+02, HO02, bullet08}. 

The dynamics of non-relativistic mixed particles is interesting and unusual. For instance, a collision of the mass eigenstate $\ket{m_h}$ with another particle can either be the elastic scattering $\ket{m_h}\to\ket{m_h}$ or the mass eigenstate conversion $\ket{m_h}\to\ket{m_l}$ (or simply the $m$-conversion $h\to l$), because of the non-diagonal elements of the flavor interaction matrix in the mass basis \cite{M10}. Let's consider $h\to l$ off a static, $\delta$-localized flavor potential with $h$ being at rest, for simplicity. The energy conservation, $m_h c^2 = m_l c^2 + m_l v^2/2$, implies that $\ket{m_l}$ gets a velocity $v=c\left[2(m_h-m_l)/m_l\right]^{1/2}$ in a random direction. Our simulations indicate that the mass-degenerate case, $m_h\simeq m_l = m$ and $\Delta m \equiv (m_h-m_l)\ll m$, fits observations the best. Thus we define the {\em `kick velocity'} parameter $v_k\equiv c\sqrt{2\Delta m/m}$, which can be used in place of the $\Delta m/m$ parameter. If $v_k$ exceeds the escape velocity from a DM halo, $v_{\rm esc}$, a part of the particle's wave-function --- the resultant $l$-eigenstate --- will escape, thus decreasing the particle's probability to be {\em in} that halo and, hence, the halo mass. This  irreversible escape of the flavor-mixed particles was called the ``quantum evaporation'' \cite{M10, M14}. The evaporation ceases if $v_{\rm esc}\gg v_k$. 

Self-interactions of two mixed DM particles is more complex and involves all 16 combinations of mass-eigenstate pairs in the input and output channels, see \citep{M14} for the full quantum mechanical analysis. The $m$-conversions in which one or two heavy eigenstates are converted, $hh\to hl$,\ $hh\to ll$ and $hl\to ll$, can lead to the quantum evaporation. Because of the energy conservation, the kinetic energy increases by $\Delta mc^2$ in processes like $hh\to hl$ and twice as much in $hh\to ll$. The reverse processes $hl\to hh$,\ $ll\to hl$ and $ll\to hh$ can also occur if kinematically allowed, i.e., if the initial kinetic energy is large enough to produce a heavy eigenstate. Finally, the elastic scattering processes $ll\to ll$,\ $hl\to hl$, $hl\to lh$, $hh\to hh$ can occur as well. 

Complete evaporation of a halo is possible depending on the $m$-conversion cross-sections, initial DM composition \citep{M14} and mixing angle, $\theta$. For simulations, we chose one such case: the maximal mixing with equal initial numbers of $h$ and $l$ eigenstates. In general, the scattering and conversion cross-sections depend on the flavor interaction strengths and $\theta$. The effect of $m$-conversions is the strongest for the maximal mixing, and 2cDM reduces to SIDM for $\theta\ll1$, see \citep{M14} for details.

{\it Implementation.} --- The physics of mixed-particle interactions was implemented in the publicly available cosmological TreePM/SPH code GADGET \citep{Springel05}. We simulated two types of DM particles representing $h$ and $l$ mass eigenstates; the total numbers of each type can change due to $m$-conversions. In the code, DM particles are interacting SPH-particles but without hydro-force acceleration. To model particles' binary interactions, we use the Monte-Carlo technique with the ``binary collision approximation'' \citep{Burkert00,Dave+01}, which is reliable for weakly collisional systems. The algorithm is as follows. For each randomly chosen projectile particle, $s_i$, a nearest neighbor is found; this is the target particle, $t_i$. For each input channel, $s_i t_i$, there are four output channels, $s_o t_o$, namely: $hh,\ hl,\ lh$ and $ll$. The probabilities of the four processes $s_i t_i\to s_o t_o$,
\beq
P_{s_i t_i\to s_o t_o}=(\rho_{t_i}/m_{t_i})\,\sigma_{s_i t_i\to s_o t_o} |{\bf v}_{t_i}-{\bf v}_{s_i}|\Delta t\ \Theta(E_{s_o t_o}) 
\label{P}
\eeq
are computed, where $\sigma_{s_i t_i\to s_o t_o}=\sigma(v)$ is the cross-section, ${\bf v}_{t_i}-{\bf v}_{s_i}$ is the relative velocity of particles in the pair, $\rho_{t_i}$ is the density of target species computed by the SPH density routine, $\Delta t$ is the iteration time-step and $\Theta(E_{s_o t_o})$ is the Heaviside function which ensures that the process is kinematically allowed (i.e., negative final kinetic energy, $E_{s_o t_o}<0$, means the process cannot occur). Whether an interaction occurs and through which channel is determined by random drawing in accordance with the computed probabilities. Kinematics of all the interactions is computed in the center of mass frame. If a scattering occurs, the particles are given random antiparallel velocities with magnitudes set by the energy-momentum conservation. If a $m$-conversion occurs, then (i) the type of one or both particles is changed, (ii) the magnitudes of the final velocities are computed with $\Delta mc^2$ given or taken, depending on the type of conversion and (iii) these velocities are assigned to the particles in antiparallel directions. If no interaction occurs, the particle velocities and types remain intact. After this, the pair is marked inactive until the next time-step. This process is repeated for all active particles at each time step. 

Our 2cDM runs have $2\times400^3=128$ million SPH-DM particles (in 2cDM, the initial numbers of $h$ and $l$ particles are equal) in the box of $50h^{-1}$~Mpc (comoving) with the force resolution scale of $3.5h^{-1}$~kpc, and the reference $\Lambda$CDM run has  $2\times640^3\approx524$ million particles and the force resolution of $2.2h^{-1}$~kpc. Our box size was optimized to be large enough to be a representative sample the universe volume, yet it provides reasonable resolution at small scales. All the runs are DM-only simulations with the standard cosmological parameters $\Omega_m=0.3,\Omega_\Lambda=0.7, \Omega_b=0$ and $h=0.7$. Initial conditions were generated using {N-GenIC} code with  $\sigma_8=0.9$ and the initial redshift $z=50$. AHF code \citep{KK09} was used to construct the halo mass function and maximum circular velocity function (MCVF), analyze halo density profiles, etc. Simulations of SIDM were done too. They fully confirm earlier studies, e.g., the inability to resolve the substructure problem, hence these results are not reported here. Numerous runs were performed to explore the range of the 2cDM model parameters $\Delta m/m$ and $\sigma/m$, to compare with the reference CDM and SIDM models and to check for numerical convergence. Here we report the most important ones. 

\begin{figure}
\includegraphics[angle = 0, width = 0.48\columnwidth]{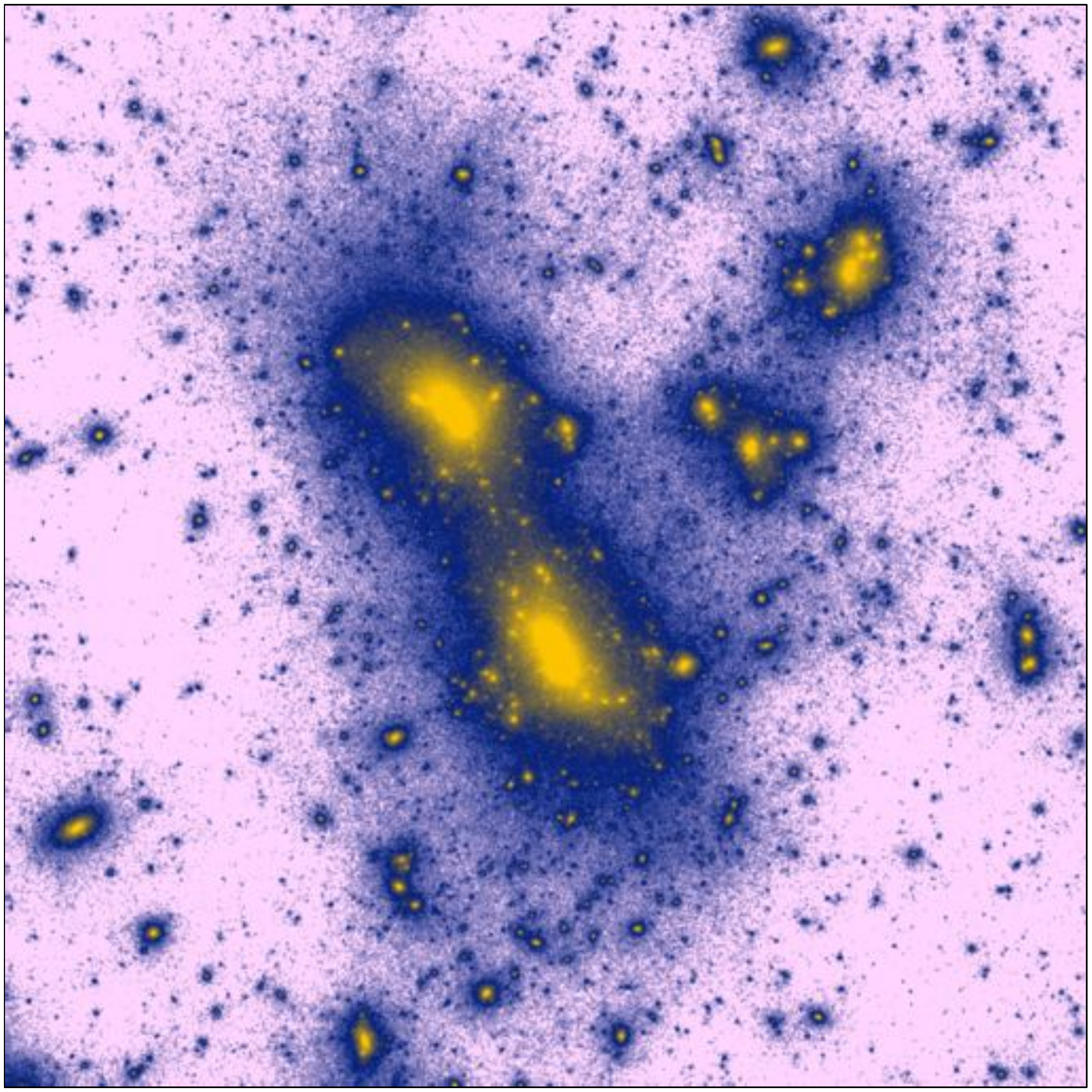} 
\includegraphics[angle = 0, width = 0.48\columnwidth]{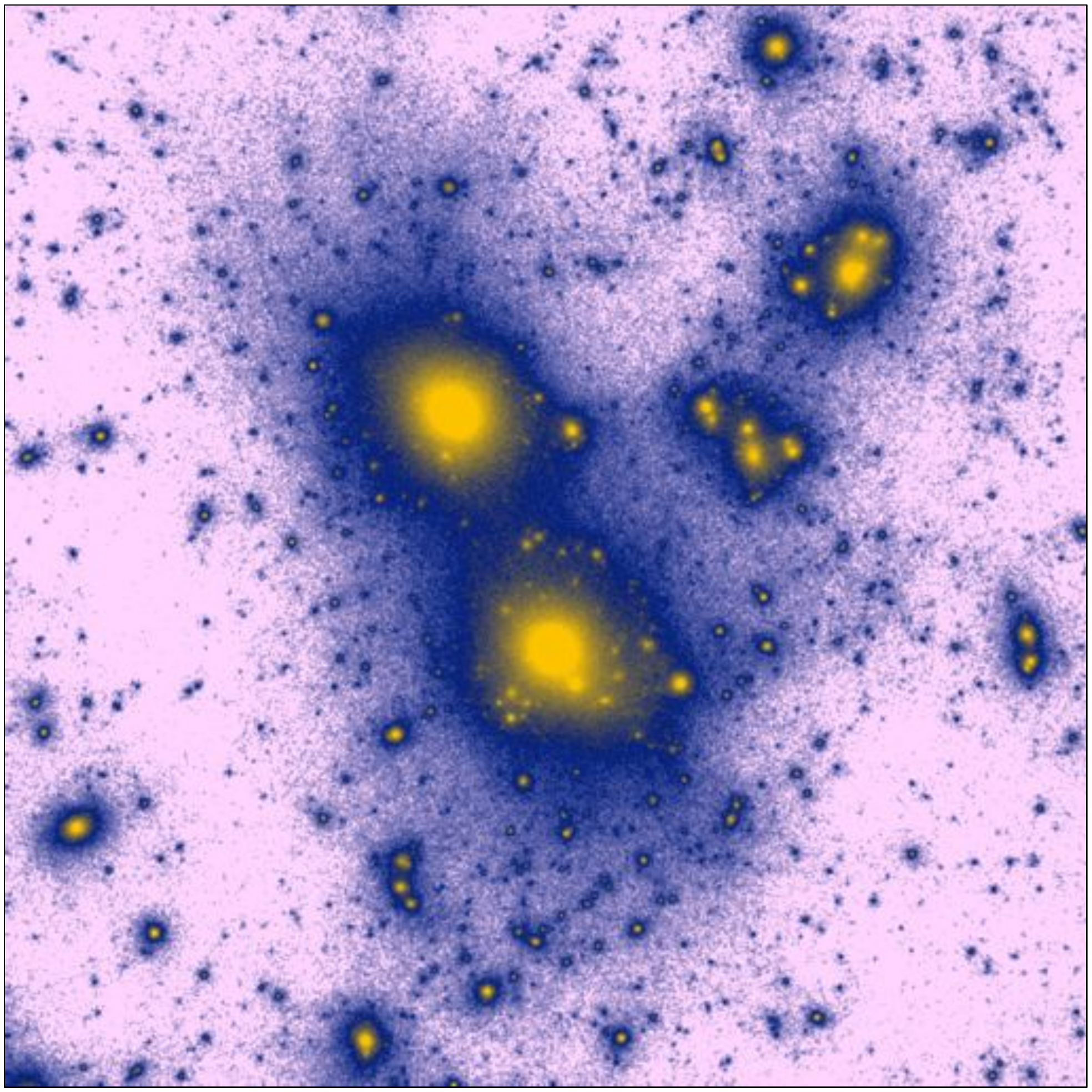}
\caption{Dark matter distribution in a region of size $5h^{-1}$~Mpc with standard $\Lambda$CDM (left panel) and 2cDM (right panel). Note the deficit of substructure (tiny yellow clumps inside large blue halos) in the 2cDM vs $\Lambda$CDM model, though large-scales remain intact.}
\label{maps-zoom}
\end{figure}
\begin{figure}
\includegraphics[angle = 0, width = 0.8\columnwidth]{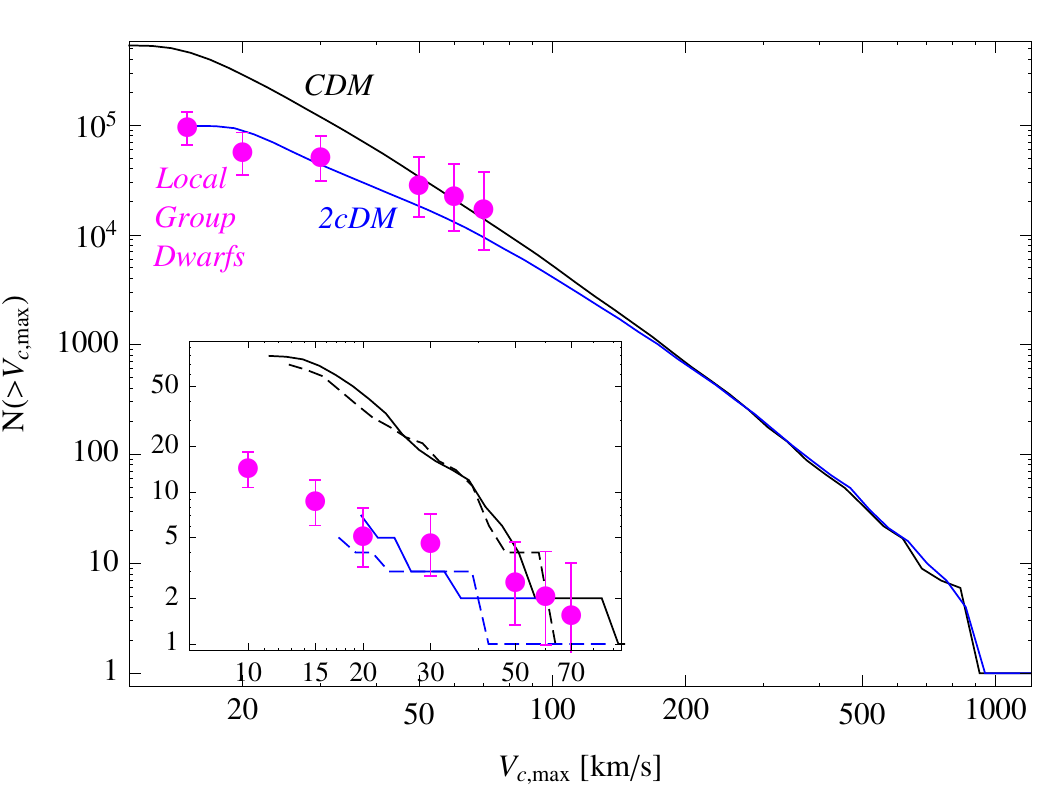}
\caption{The MCVF in the entire simulation box for $\Lambda$CDM (black curve) and 2cDM (blue curve). The 2cDM model provides an excellent fit to the rescaled Local Group data \citep{Kravtsov10,Klypin+99} (magenta points).  The inset compares the MCVFs for the substructure within 571~kpc around two individual Milky Way-like halos, as in \citep{Klypin+99}, against original (non-rescaled) data.}
\label{keyfig}
\end{figure}

{\it Results.} --- Simulations with the large mass difference $m_h\ge m_l$ (not presented here) grossly disagree with the observational data, so this case is not considered further. Hence, because of  $\Delta m/m\ll1$, the mass segregation of heavier species toward the halo center is negligible. 

The DM maps in a zoomed-in region of 5~Mpc across at $z=0$ for the 2cDM and $\Lambda$CDM models are presented in Fig. \ref{maps-zoom}. One sees fewer subhalos in the 2cDM case. The parameters are $\Delta m/m\simeq10^{-8}$, which corresponds to $v_k=50$~km/s, and $\sigma/m=0.75$~cm$^2$/g at $v_0\sim v_k$, which is fully consistent with observational constraints on the SIDM cross-section \citep{HO02, Colin+02, +Shapiro+05, bullet08, A-H+09, F+10, LW11, VZL12}. For these values, the 2cDM MCVF matches the Local Group data the best, as shown in Fig. \ref{keyfig}. This figure shows the number of halos with the maximum circular velocity above a certain value, $N(>V_{c,{\rm max}})$ versus $V_{c,{\rm max}}$,  for 2cDM and $\Lambda$CDM; the data points are from \citep{Kravtsov10,Klypin+99}. The amount of substructure is volume-dependent, so we appropriately rescaled the data points to reproduce the results of Refs. \citep{Klypin+99,Kravtsov10} using the MCVF from our $\Lambda$CDM simulation; the procedure is legitimate for a scale-free ergodic distribution of DM structure. However, no data rescaling is done for the substructure MCVFs of two individual Milky Way-like halos shown in the inset. In both cases, the agreement with 2cDM is much better than with $\Lambda$CDM. 

The simulations show that $v_k$ uniquely determines the position of the break in the MCVF, $V_{c,{\rm max}}^\text{break}\simeq v_k$, whereas $\sigma/m$ determines the slope below the break. By comparing simulations with observational data, we determined $v_k$ (and consequently $\Delta m/m$) to be around $\sim 50-70$~km/s. Interestingly, a similar value of a characteristic velocity $\lesssim100$~km/s was found in another independent analysis of survey data \citep{Zwaan+10}. The `best fit' cross-section is $\sigma/m\sim0.75$~cm$^2$/g at $v_k$ but values a factor of two smaller or larger are acceptable too. The halo mass function exhibits the even sharper break at $M\simeq 10^{10}M_\odot$. Thus, the overall suppression of the abundance of dwarf halos resolves the substructure problem.

\rem{
\begin{figure*}
\includegraphics[angle = 0, width = 0.5\columnwidth]{profilesCDM_} 
\includegraphics[angle = 0, width = 0.5\columnwidth]{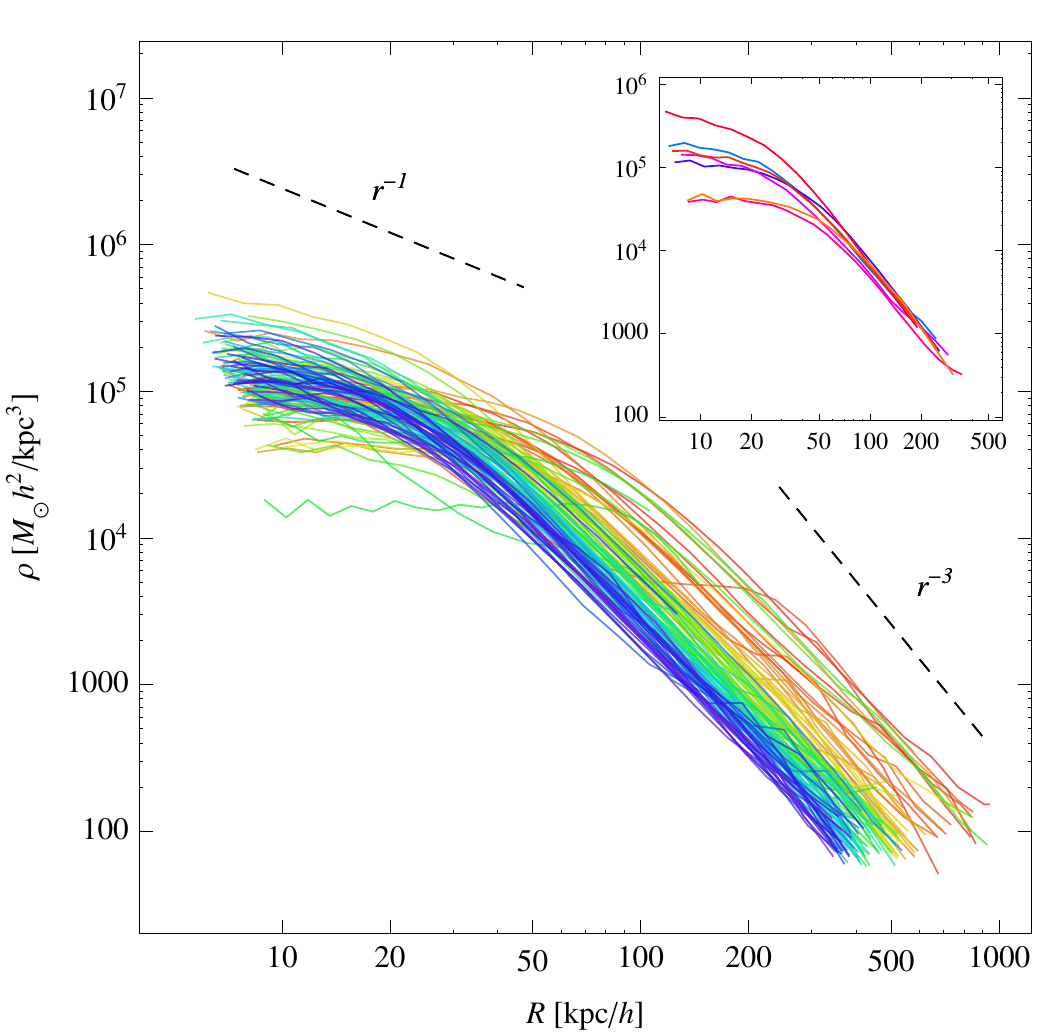}
\includegraphics[angle = 0, width = 0.5\columnwidth]{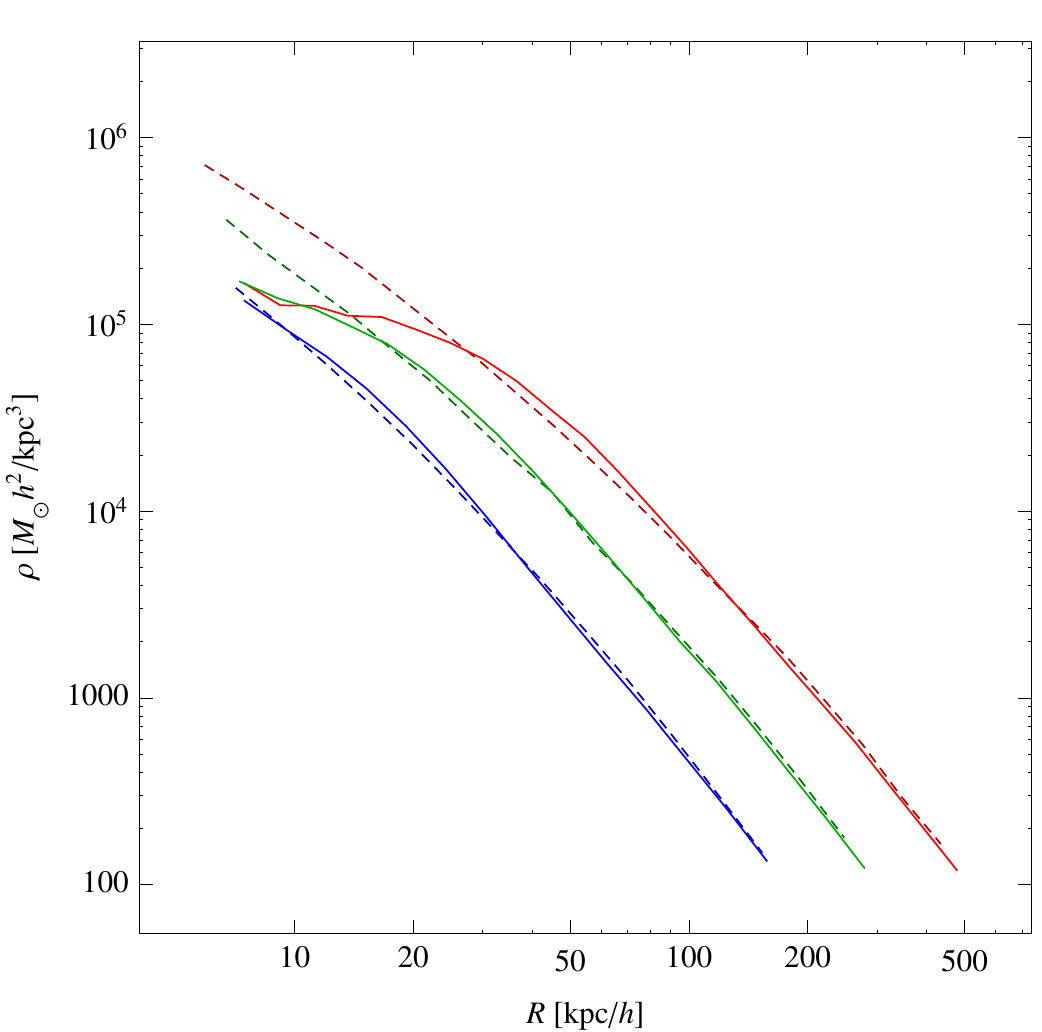}
\caption{Density profiles of 120 well-resolved dark halos in $\Lambda$CDM (left panel) and 2cDM (middle panel) models. 2cDM exhibits flatter profiles. The profiles are color-coded by the halo mass: red -- most massive, blue -- less massive. The right panel shows the averaged CDM (dashed) and 2cDM (solid) profiles obtained by stacking the profiles within a narrow, $\sim30\%$, mass range around $2\times10^{13}M_\odot$ (red), $4\times10^{12}M_\odot$ (green) and $8\times10^{11}M_\odot$ (blue). The inset shows six 2cDM individual halo profiles with masses between $(2-1.7)\times10^{13}M_\odot$ for 2cDM. The large central density variance makes the stacked 2cDM profiles unrepresentative.}
\label{prof}
\end{figure*}
}

\begin{figure}
\includegraphics[angle = 0, width = 0.49\columnwidth]{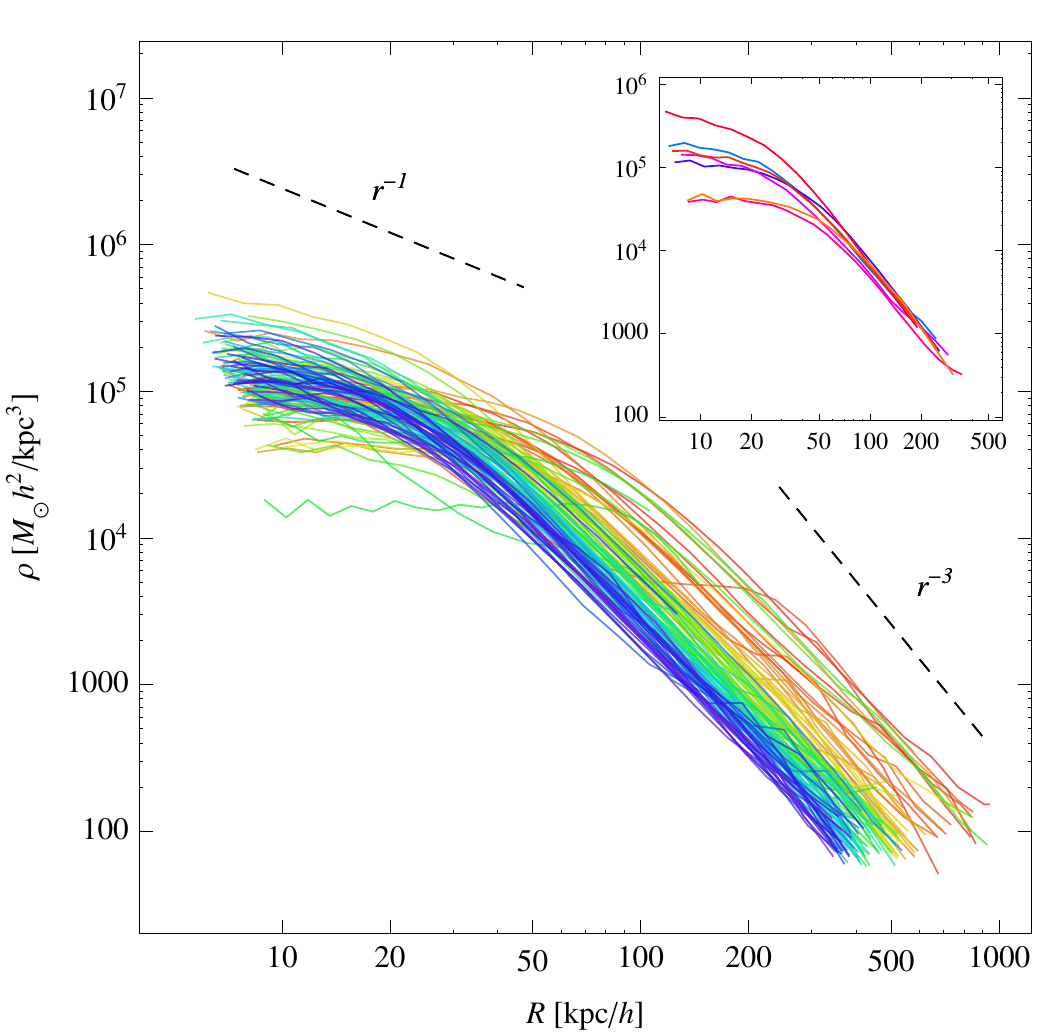}
\includegraphics[angle = 0, width = 0.49\columnwidth]{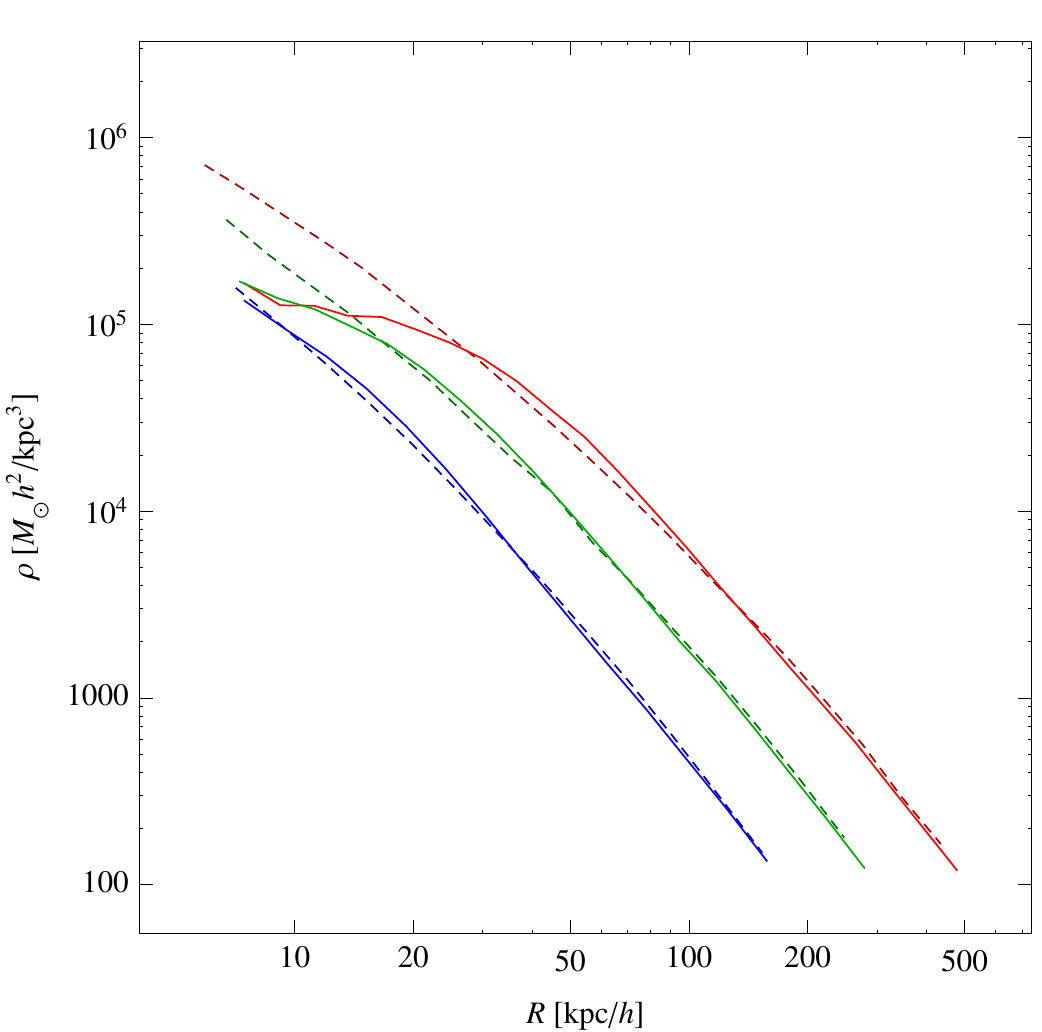}
\caption{Left panel: density profiles of 120 well-resolved 2cDM dark halos; they are flatter than $1/r$. The profiles are color-coded by the halo mass: red -- most massive, blue -- less massive. Right panel: averaged CDM (dashed) and 2cDM (solid) profiles obtained by stacking the profiles within a narrow, $\sim30\%$, mass range around $2\times10^{13}M_\odot$ (red), $4\times10^{12}M_\odot$ (green) and $8\times10^{11}M_\odot$ (blue). The inset shows six 2cDM individual halo profiles with masses between $(2-1.7)\times10^{13}M_\odot$ for 2cDM. The large central density variance makes the stacked 2cDM profiles unrepresentative.}
\label{prof}
\end{figure}
\begin{figure}
\includegraphics[angle = 0, width = 0.75\columnwidth]{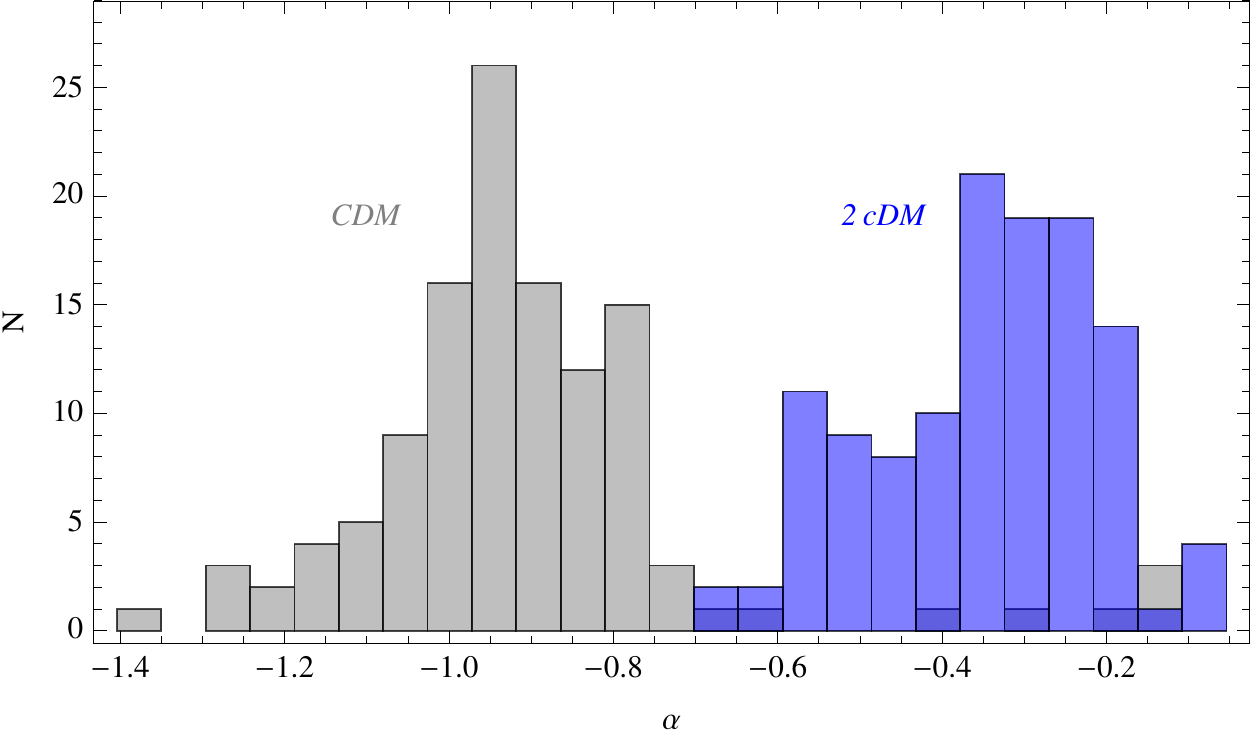}
\caption{Histograms of the slopes of the inner density profiles of the halos shown in Fig. \ref{prof}. Whereas the CDM profiles show a cusp $r^\alpha$ with $\alpha\sim -0.8 \ldots -1$ consistent with earlier studies, the 2cDM profiles are much shallower: $\alpha\sim-0.2\ldots-0.6$.}
\label{alphas}
\end{figure}

Fig. \ref{prof} shows 120 well-resolved halo density profiles for $\Lambda$CDM and 2cDM. The profiles are trustworthy everywhere because their inner parts were truncated according to the numerical binary collision criterion \citep{power+03}. The $\Lambda$CDM profiles agree with the NFW profile. In contrast, the 2cDM inner profiles are shallower and less centrally concentrated. Although the stacked profiles are less noisy, the inset shows that individual profiles exhibit an order-of-magnitude variance in their central densities due to different formation histories. The lack of a common central density scale agrees with observations \citep{KdN+10}, which were originally argued against SIDM. 

The softening of cusps is also seen from Fig. \ref{alphas}. Here, the effective power-law index is obtained by fitting the individual profiles with the function $\rho=\rho_0\,r^\alpha(1+r/r_c)^\beta$ and then evaluating $\alpha$ at $r=7$~kpc/$h$. The distribution of the slopes ranges within $\alpha\simeq-0.8\ldots-1$ for CDM indicating a cusp and within $\alpha\simeq-0.2\ldots-0.6$ for 2cDM, which thus explains the core/cusp and, likely, too-big-to-fail problems. Importantly, the density profiles and core sizes of massive halos are mostly sensitive to $\sigma/m$, whereas $v_k$ plays little role, if any. The profiles of the halos with  $M\lesssim10^{10}M_\odot$ may depend on $v_k$, which should be explored with dedicated high-resolution simulations. We stress that SIDM runs show softened cusps but no substructure suppression -- the MCVF is a scale-invariant power-law -- due to the lack of a physical parameter, such as $v_k$ in 2cDM, which can set the break scale.

{\it Implications.} --- (i) Cosmology with at least two flavor-mixed mass-degenerate, $\Delta m/m\sim10^{-8}$, species can naturally resolve cosmological problems at small scales without invoking new or exotic physics. In contrast, single-species and/or non-mixed  candidates, and non-degenerate multi-component models are disfavored.

(ii) 2cDM agrees with observations within a range of the velocity-dependent $\sigma/m$ allowed for SIDM \citep{HO02, Colin+02, +Shapiro+05, bullet08, A-H+09, F+10, LW11, VZL12}. The constraints are tight: if $\sigma$ is too small, then it is cosmologically uninteresting, if it is too large, then the cusps will be enhanced due to the gravithermal collapse of halos. This fine tuning, rephrased as the ``Why now?'' question is a caveat of 2cDM. However, SIDM and dark energy/cosmological constant face the same problem. 

(iii) Our model does not change the linear power spectrum, unlike WDM; all changes occur in the nonlinear stage. The quantum evaporation proceeds slowly over the Hubble time. We can speculate that the gas metal-enriched by the stars in dwarf spheroidals should gradually become unbound from the weakening gravitational potential of the halos and enrich the intergalactic medium with metals, resembling the effect of supernova/winds. Since not all small halos are evaporated by $z=0$, the residual substructure can be responsible for the flux anomalies in gravitational lensing observations. 

(iv) Our simulations can formally describe any multi-component DM where transformations of species are allowed. However, these models face a severe problem: Why have the heavy (e.g., `excited', etc.) particles survived in the early universe, but convert to lighter (or `ground-state') species now, when the density is much smaller? The flavor-mixed 2cDM model does not have this problem, because the $m$-conversion cross-section in the flat space-time is suppressed by $(\Delta m/m)^4\sim10^{-32}$ over it's current value \citep{M14} and becomes large only during the structure formation, when mass eigenstates separate. 

(v) The 2cDM theory is testable with direct detection experiments. Indeed, DM is a collection of $h$ and $l$ eigenstates, which can convert into one another in interactions with normal matter in a detector. These conversions should result in the energy `mismatch' of $\sim \pm\Delta mc^2$, i.e., the events will look like {\em inelastic} collisions: ``exothermic'' and ``endothermic". Particularly, the down-conversions $h\to l$, which are always kinematically allowed, can look like ``exothermic" interactions. In contrast, the $l\to h$ up-conversions can occur only if the kinetic energy exceeds a threshold. Hence, the $l\to h$ rate can exhibit a stronger annual modulation.  Next, we can also speculate that if DAMA and CoGeNT anomalies are due to inelastic effects with $\Delta m\sim$~keV, then the DM mass is $m\sim10^8\Delta m\sim 10^2$~GeV, which is close to that inferred from the GeV excess in {\em Fermi}-LAT data \citep{Weniger12,KP14}. Finally, we also suggest that the use of different targets (e.g., Ne, Ar) in the experiments may strongly affect the recoil signal strength because of possible different flavor couplings to the DM species, whose flavor composition is unknown. 

(vi) 2cDM can be tested in indirect detection experiments. For instance, the direct DM annihilation into two photons results in a line triplet corresponding to the annihilations in $h+h,\ h+l$ and $l+l$ channels. Thus, the DM annihilation line can be a triplet at $E=mc^2$ spit by $\Delta E=\frac{1}{2}\Delta mc^2$ and with different line strengths.

The author is grateful to Lars Hernquist, Avi Loeb, Ramesh Narayan, Lyman Page, Sergei Shandarin, Mark Vogelsberger for discussions and suggestions. This work was supported in part by the Institute for Theory and Computation at Harvard University and by DOE grant  DE-FG02-07ER54940, NSF grant AST-1209665 and XSEDE grants AST110024 and AST110056. The simulations utilized XSEDE high-performance computing systems {\it Trestles} (SDSC) and  {\it Ranger} (TACC).

\end{document}